\documentclass[12pt]{article}
\usepackage[a4paper,left=25mm,top=25mm,textwidth=160mm,textheight=249.7mm]{geometry}
\usepackage{authblk}
\usepackage{natbib}

\usepackage{graphicx}	
\usepackage{amsmath}	
\usepackage[breaklinks=true]{hyperref}
\usepackage{listings}
\usepackage{xcolor}
\usepackage{ulem}

\usepackage{multicol}

\providecommand{\keywords}[1]
{
  \small	
  \textbf{\textit{Keywords ---}} #1
}

\usepackage{tikz}

\definecolor{lime}{HTML}{A6CE39}
\DeclareRobustCommand{\orcidicon}{%
	\begin{tikzpicture}
	\draw[lime, fill=lime] (0,0) 
	circle [radius=0.16] 
	node[white] {{\fontfamily{qag}\selectfont \tiny ID}};
	\draw[white, fill=white] (-0.0625,0.095) 
	circle [radius=0.007];
	\end{tikzpicture}
	\hspace{-2mm}
}

\foreach \x in {A, ..., Z}{%
	\expandafter\xdef\csname orcid\x\endcsname{\noexpand\href{https://orcid.org/\csname orcidauthor\x\endcsname}{\noexpand\orcidicon}}
}

\definecolor{codegreen}{rgb}{0,0.6,0}
\definecolor{codegray}{rgb}{0.5,0.5,0.5}
\definecolor{codepurple}{rgb}{0.58,0,0.82}
\definecolor{backcolour}{rgb}{0.95,0.95,0.92}

\lstdefinestyle{mystyle}{
  backgroundcolor=\color{backcolour},   commentstyle=\color{codegreen},
  keywordstyle=\color{magenta},
  numberstyle=\tiny\color{codegray},
  stringstyle=\color{codepurple},
  basicstyle=\ttfamily\footnotesize,
  breakatwhitespace=false,         
  breaklines=true,                 
  captionpos=b,                    
  keepspaces=true,                 
  numbers=left,                    
  numbersep=5pt,                  
  showspaces=false,                
  showstringspaces=false,
  showtabs=false,                  
  tabsize=2
}

\title{Pytearcat: PYthon TEnsor AlgebRa calCulATor\\
\Large{A python package for general relativity and tensor calculus}}

\author[1,2]{M. San Martín\orcidA{}}
\author[1,2]{J. Sureda\orcidB{}}
\affil[1]{\normalsize{Instituto de Astrofísica, Pontificia Universidad Católica de Chile, Vicuña Mackenna 4860, Santiago, Chile}}
\affil[2]{\normalsize{Centro de Astro-Ingeniería, Pontificia Universidad Católica de Chile, Vicuña Mackenna 4860, Santiago, Chile}}

\begin{document}

\maketitle

\begin{abstract}
This paper introduces the first release of Pytearcat, a Python package developed to compute tensor algebra operations in the context of theoretical physics, for instance, in general relativity. Given that working with tensors can become a complex task, people often rely on computational tools to perform tensor calculations. We aim to build a tensor calculator based on Python, which benefits from being free and easy to use. Pytearcat syntax resembles the usual physics notation for tensor calculus, such as the Einstein notation for index contraction. This version allows the user to perform many tensor operations, including derivatives and series expansions, along with routines to obtain the typical General Relativity tensors. A particular concern was put in the execution times, leading to incorporate an alternative core for the symbolic calculations, enabling to reach much faster execution times. The syntax and the versatility of Pytearcat are the most important features of this package, where the latter can be used to extend Pytearcat to other areas of theoretical physics.
\end{abstract}



\section{Introduction}

As physics and theoretical physics develop, the necessity to perform complex calculations also raises as new theoretical models appear. At some point, these tasks become very hard or unpractical to do by hand. Thus, the use of computational tools becomes a necessity. In General Relativity (GR), this becomes important since tensor operations involve the calculation of many symbolic calculations at once. For this reason, many symbolic calculators work with tensor calculus, such as GRTensorIII\footnote{The details about this Maple Package can be found on its web page \href{http://grtensor.phy.queensu.ca}{(grtensor.phy.queensu.ca/)} and its GitHub repository \href{https://github.com/grtensor/grtensor}{(github.com/grtensor/grtensor})} or Einsteinpy \citep{Einsteinpy:2020}. 

There are differences and advantages of using one or other available option that will depend mainly on the user's objective. Nevertheless, some of these options work on commercial software (Maple or Mathematica), which becomes a barrier to the users. Furthermore, some other problems are related to the fact that some packages are created to perform specific tasks, making the work difficult and tedious. For instance, other programs that compute the typical GR tensors, do not allow to define new tensors and operate them. Then it becomes nearly impossible to calculate other things, such as perturbation theory with a non-standard metric in modified gravity theories; For example, some tensor definitions involved in non-standard theories, can have long and complicated expressions (see, for instance, the equations in \citep{SanMartin:2021}), where Pytearcat can be very useful in order to obtain and work with these kind of equations.

This paper presents Pytearcat: PYthon TEnsor AlgebRa calCulATor. An open-source Python package (under the GNU general public license version 3) created to work with general tensor operations, either in the field of GR or others that need to use tensor calculus. It provides the basic GR tensors built in the package and uses a standard syntax for the Einstein notation. 

The paper has the following structure: In section \ref{sec:structure} we describe the structure of the package, indicating where the main functionalities locate; In section \ref{sec:definitions} we explain the main definitions used in the package, related to the GR framework; Later, in section \ref{sec:tensor} we show the general tensor operations and their usage within the package, along with some features present in the package; Finally, in section \ref{sec:execution} we compare the execution times of the package using different symbolic operators and other packages available; we end the paper with the conclusions of this work and ideas for the future development of Pytearcat.

\section{Package structure}\label{sec:structure}

\begin{figure}
    \centering
    \includegraphics[width=0.8\textwidth]{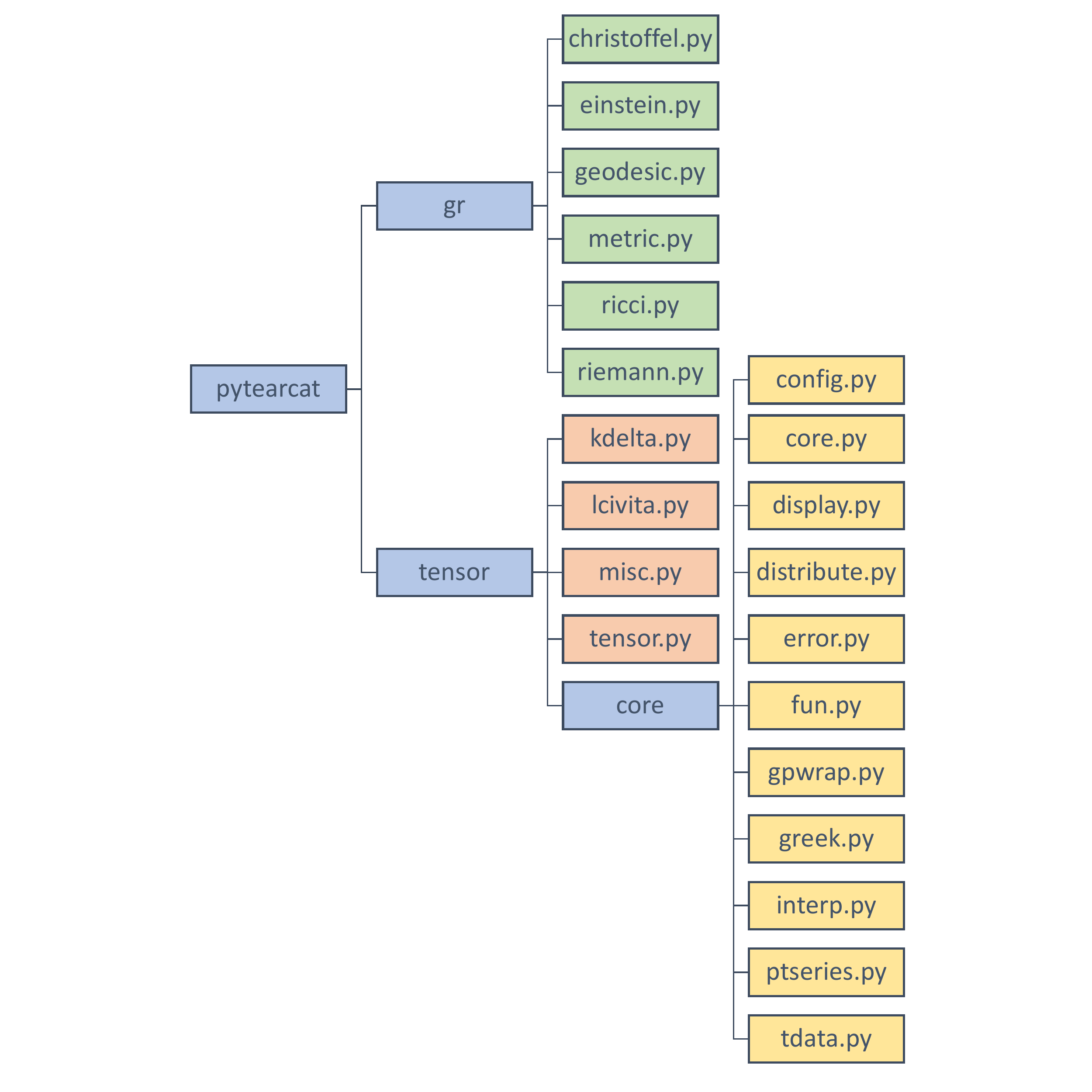}
    \caption{Structure of Pytearcat. The blue indicates the packages and sub-packages. The green boxes are the modules that are useful to work in GR. The salmon shows modules that contains useful routines for the user. The yellow boxes are a deeper part of the code, and they contain routines that the other modules require, but the user would not call them.}
    \label{fig:tree}
\end{figure}

We present a tree diagram summarising the Pytearcat package in Figure \ref{fig:tree}. Inside the Pytearcat package, there are two sub-packages named \textit{gr} and \textit{tensor}. Inside the first one, there are six modules related to GR expressions (indicated as green boxes in Figure \ref{fig:tree}). These libraries allow calculating quantities that are very common in GR, such as the Christoffel symbols (first and the second kind, \textit{christoffel.py}), the Ricci tensor and the Ricci scalar (\textit{ricci.py}), the Riemann tensor (\textit{riemann.py}) and the Einstein tensor (\textit{einstein.py}). Also, there is a module to define the metric (\textit{metric.py}) and another to calculate the geodesics (\textit{geodesic.py}).
The second sub-package named \textit{tensor} contains modules that allow to define tensors and operate with them. The modules inside this folder are coloured as salmon in Figure \ref{fig:tree}. The \textit{misc.py} module contains functions that allow defining symbolic functions, variables, and constants. It also contains other functions to work with series expansions and to simplify expressions. The \textit{kdelta.py} and \textit{lcivita.py} modules contain the data classes which define the Kronecker Delta symbol and the Levi-Civita symbol, respectively. 
The \textit{tensor.py} module contains the code related with the class \texttt{tensor} and many functions that are useful to define a tensor, operate tensors, recognise the contravariant and covariant indices, lower and raise indices and expand a tensor like a series up to a specific order. Inside this sub-package, there is another one named \textit{core} which contains essential information that the program needs to operate tensors. All the functions required by the user are located at the top level of the package. 

The Pytearcat core works by default with Sympy \citep{sympy} and, optionally, with Giacpy\footnote{Giac/Xcas, Bernard Parisse and Renée De Graeve, version 1.7.0 (2021), \href{https://www-fourier.univ-grenoble-alpes.fr/~parisse/giac.html}{\nolinkurl{www-fourier.univ-grenoble-alpes.fr/~parisse/giac.html}}}, depending on the installation or the environment. These are two symbolic calculators with advantages and disadvantages. By default, the symbolic calculations are carried out by Sympy, meaning that each component of a tensor corresponds to a Sympy object. This core is beneficial for combining calculations with Sympy variables, but it can be very inefficient in performing some tasks. For this reason, we included the possibility of working with Giacpy for the symbolic calculations. This package is a Python wrapper of a symbolic calculator programmed in C. Giacpy performs better than Sympy, but its documentation and information are hard to find. We are aware that using two different cores for the packages may be confusing, and we tried to use Symengine\footnote{The details of this package are available on its web page \href{https://symengine.org}{symengine.org}.} as the main core. Symengine would allow the user to benefit from the Sympy variables with a fast engine without worrying about compatibility issues. However, this could not be possible since some features that we think would be important do not work correctly in Symengine, and we prefer to wait until these problems are solved. One of the advantages of using these cores are that functions and variables defined with them are compatible with Pytearcat's objects. For example, if the users are working with the Sympy core, they can assign a trigonometric function (\texttt{sympy.cos(x)}) to a tensor component without compatibility problems.

Finally, since Pytearcat is a package to perform tensor calculus, it needs to run on an environment that shows readable outputs. Jupyter Notebooks is the perfect choice for this as it can render \LaTeX $ $ outputs, show information about the current processes, and is widely used in astronomy and astrophysics with a user-friendly interface. Also, the output of a Jupyter cell is an object that can be manipulated as desired using the corresponding methods from either Pytearcat or the core. 

\subsection{Installation}

The source code is on its GitHub repository \href{https://github.com/pytearcat/pytearcat}{(github.com/pytearcat/pytearcat)}, and as a Python package, it can be installed through pip as shown in listing \ref{listing:pip_wo_gp}.

\begin{lstlisting}[caption= {Installing Pytearcat in a Python environment.},label = listing:pip_wo_gp]
    pip install pytearcat
\end{lstlisting}

To use the Giacpy core in Pytearcat, the user must explicitly indicate during the package installation that the installer must include the Giacpy module in the process. This is shown in listing \ref{listing:pip_w_gp}.

\begin{lstlisting}[caption= {Installing Pytearcat with giacpy in a Python environment.},label = listing:pip_w_gp]
    pip install pytearcat[giapy]
\end{lstlisting}

For now, Pytearcat is only compatible with Python 3.7.

If you find any problem with Pytearcat you can report the issue on \href{https://github.com/pytearcat/pytearcat/issues}{GitHub}.

\section{Basic definitions}\label{sec:definitions}

\subsection{What is a tensor?}

In this package, we have adopted notations and definitions commonly used in physics and astrophysics \citep{book:Carroll:2014}. In this context, we will say that an $n$th-rank tensor in $m$-dimensional space is a mathematical object with $n$ indices and $m^n$ components that follow specific transformation rules. Each of these indices can be evaluated in one of the $m$ coordinates that describe the space-time.

A general tensor of rank $n$ can be of mixed type, and it is denoted as $(r,s)$ where $n=r+s$. In this case, $r$ is called "contravariant" (upper) indices and $s$ "covariant" (lower) indices. The order of the indices is important. A tensor with rank 0 is called scalar, but the program does not allow these cases. If the user wants to work with a scalar, it must be defined as a constant, variable, or function. The tensors which transform like a 1-rank tensor are vectors, and the 2-rank are called matrices. In our notation, a vector $v$ is written as $v_i$ where $i = 1,...,m$. This is different from $v^i$, and the display of the object also is different. In Pytearcat the objects are printed as follows:

\begin{equation}
v_i = (a,b,c,d,...)
\end{equation}

and

\begin{equation}
v^i = \begin{pmatrix}
a\\
b\\
c\\
d\\
\vdots\\
\end{pmatrix}
\end{equation}

Then, in general, we are going to write a tensor with physics notation, for example, ${B}^{i}{}_l{}^j{}^k{}_m{}$. Note that the indices are placed in the corresponding position, where $i$ is the first index (contravariant), $l$ is the second index (covariant), $j$ and $k$ are the third and fourth indices (contravariant), $m$ is the fifth index (covariant).

According to our notation, we can create a tensor object as an instance of the \texttt{Tensor} class. This instance has all the object's information and has all the indices combination that a tensor of rank $m$ has. For example, an instance called $A$ of the \texttt{Tensor} class with rank 3, has three indices and contain all the combinations, in other words the object $A$ contains the following elements: $A{}_i{}_j{}_k$, $A{}^i{}_j{}_k$, $A{}_i{}^j{}_k$,$A{}_i{}_j{}^k$,$A{}^i{}^j{}_k$,$A{}^i{}_j{}^k$,$A{}_i{}^j{}^k$ and $A{}^i{}^j{}^k$. We are going to detail how to define and work with tensors in the section \ref{sec:tensor}. In the following subsections, we show the basics applications of Pytearcat.

\subsection{The metric tensor}

In Pytearcat is essential to define the metric tensor. In this step, the user defines the dimension of the space-time and which coordinate is associated with time. To work with a \texttt{Tensor} object, we must define a metric tensor to raise and lower an index. In our physics notation, we denote the $g$ metric tensor as a 2-rank tensor which is symmetric ($g{}_i{}_j = g{}_j{}_i$). In physics, this object is commonly related to the extension of the Pythagorean length in curved spaces, where the line element $ds$ measures the infinitesimal length where

\begin{equation}
  ds^2=g_{00}dx^0dx^0+g_{01}dx^0dx^1+g_{10}dx^{1}dx^{0}+...+g_{nn}dx^ndx^n.
  \label{eq:lineelement}
\end{equation}
 
 $g{}_i{}_j$ appears in front of every combination of two differentials of the coordinates that describe the space. An $n$ dimensional space is described by $n$ independent coordinates given by $x^0,...,x^{n-1}$, and the corresponding differential forms are $dx^0,...,dx^{n-1}$. The metric tensor is defined by the coefficients that appear in front of every $dx^idx^j$ combination of the line element in equation \eqref{eq:lineelement}.

In Pytearcat, we need to define the coordinates before working. These coordinates are initialised calling the function \texttt{coords} directly from Pytearcat. All the functions that the user would use can be called directly from Pytearcat. Other functions are useful before defining the metric. For example, we can define a function calling \texttt{fun} and passing it two arguments. They must be strings where the first is the object's name, and the second contains the variables that have to be separated with commas. For example, if we want to define a function $f = f(t,x)$ we have to write \texttt{f = pt.fun('f','t,x')}. If the user wants to re-define the function, the attribute \texttt{overwrite = True} must be included. In our simplified example, we work with the Friedman-Lemaître-Robertson-Walker metric (FLRW) given by 

\begin{equation}
 \mathrm {d} s^{2}=-\mathrm {d} t^{2}+{a(t)}^{2}\left({\frac {\mathrm {d} r^{2}}{1-kr^{2}}}+r^{2}\mathrm {d} \theta ^{2}+r^{2}\sin ^{2}\theta \,\mathrm {d} \phi ^{2}\right).
\end{equation}

\begin{figure}
    \centering
    \includegraphics[width=0.8\textwidth]{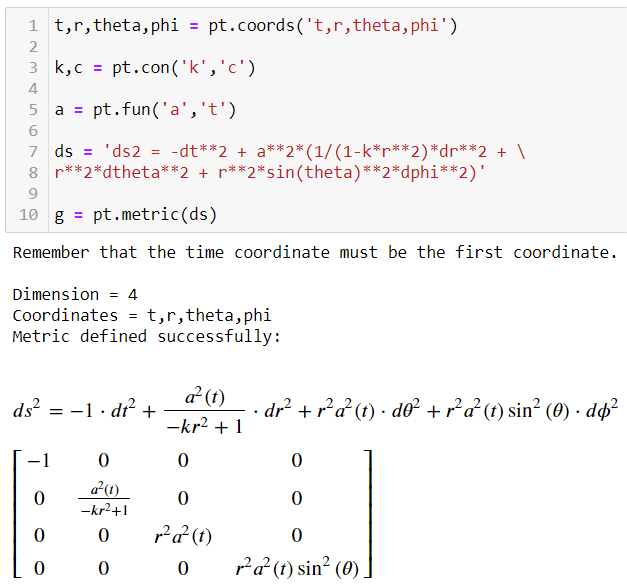}
    \caption{This example shows how to define the FLRW metric, defining the coordinates, the constants, the scale factor as a function of time, and finally calling the function \texttt{metric} which receives a string of the line element. It also shows the output of the execution of the cell, displaying the line element and the metric tensor.}
    \label{fig:metric}
\end{figure}

First, we have to define the manifold coordinates, and then we can define the function $a(t)$ known as scale factor, and finally, we can call the function \texttt{metric} to define the metric. This procedure is shown in Figure \ref{fig:metric}. Also, it is important to define the curvature $k$, which is a constant. This is defined through the function \texttt{con} as it is shown in Figure \ref{fig:metric}.

The function \texttt{metric} always print the line element and the metric tensor $g_{\mu\nu}$. This function also shows some important parameters such as the dimension and the coordinates. 

We stress that Pytearcat only works with symmetric metrics and only support torsion-free connections.

With the coordinates and the metric, the program can calculate all the standard tensors. First, we will introduce the standard tensors (associated with GR), and then we will show how to work with arbitrary tensors.

\subsection{General relativity tensors}\label{sec:GR-Tensor}

There are five mathematical objects that are essential to work in GR. These are the Christoffel symbols, the Riemann tensor, the Ricci tensor, the Ricci scalar and the Einstein tensor.

We have to clarify that a \texttt{Tensor} object (an instance of the \texttt{Tensor} class) does not distinguish if the defined object is a tensor (as defined in physics) or is a mathematical object with upper and lower indices. By default, in Pytearcat, a \texttt{Tensor} is instantiated with all its components as \texttt{NAN} values until they are calculated or get values assigned to them. The definition adopted in Pytearcat creates much flexibility in defining, assigning, and operating them. An object with indices can be stored in an instance of the \texttt{Tensor} class, and its other indices will be obtained using the classic rules of raising and lowering indices. For example, a physically meaningful object $B{}^\mu{}_\nu$ can be assigned to an instance of the \texttt{Tensor} class $A$, in the component $A{}^\mu{}_\nu$. If the user asks for the program to automatically raise and lower indices, through the \texttt{complete} class-method, Pytearcat will automatically treat the object as a tensor, e.g., $B{}^\mu {}^\nu=g{}^\nu{}^\alpha B{}^\mu{}_\alpha $.
Finally, the metric, the Ricci tensor, the Riemann tensor, and the Einstein tensor are \texttt{Tensors}. Meanwhile, the Christoffel of the first and second types are instances of the \texttt{ChristoffelClass}. Any mathematical operation between objects from Pytearcat will return data that can be stored inside an instance of the \texttt{Tensor} class. We show examples of these cases in section \ref{sec:tensor}.
\subsubsection{Christoffel symbols}

The Christoffel symbol of the first kind is

\begin{equation}
\Gamma _{\gamma\alpha\beta}={\frac {1}{2}}\left({\frac {\partial g_{\gamma\alpha}}{\partial x^{\beta}}}+{\frac {\partial g_{\gamma\beta}}{\partial x^{\alpha}}}-{\frac {\partial g_{\alpha\beta}}{\partial x^{\gamma}}}\right),
\label{eq: Chr_first}
\end{equation}

and the Christoffel symbol of the second kind is

\begin{equation}
{\Gamma^{\gamma}}_{\alpha \beta}={\frac {1}{2}}g^{\gamma \lambda}\left({\frac {\partial g_{\lambda\alpha}}{\partial x^{\beta}}}+{\frac {\partial g_{\lambda\beta}}{\partial x^{\alpha}}}-{\frac {\partial g_{\alpha \beta}}{\partial x^{\lambda}}}\right).
\label{eq: Chr_second}
\end{equation}

\begin{figure}
    \centering
    \includegraphics[width=0.8\textwidth]{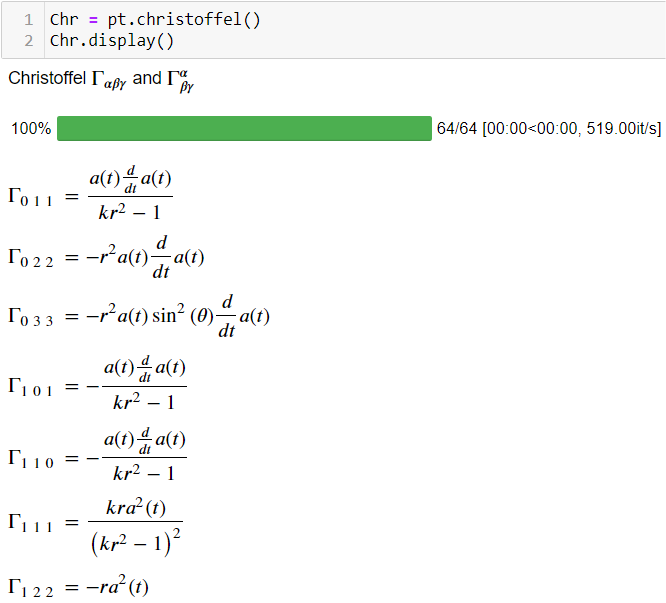}
    \caption{Example of the calculation of the Christoffel symbols and the \texttt{display} method, together with the respective output. In this screenshot we do not show all the Christoffel components.}
    \label{fig:Christoffel}
\end{figure}

We can calculate them calling the function \texttt{christoffel} as we show in Figure \ref{fig:Christoffel}. This function uses the defined metric to calculate the Christoffel symbols through Eq. \eqref{eq: Chr_first} and \eqref{eq: Chr_second}. In this example this corresponds to the metric defined in Figure \ref{fig:metric}. 

In Pytearcat, tensor operations that depend directly on the metric use the underlying metric and the corresponding Christoffel symbols. This is to avoid confusion regarding which metric tensor is being used to raise or lower indices of a tensor, or which Christoffel symbols are used to calculate covariant derivatives. Keep in mind that at any point the user is able to redefine the metric and therefore to recalculate everything with a different metric.

In Figure \ref{fig:Christoffel}, the execution of the first line will calculate the two Christoffel symbols (the first and the second kind). \texttt{christoffel} can receive two Boolean arguments which are defined by default as \texttt{First\_kind=True} and \texttt{Second\_kind=True}. If we want to calculate only one kind, we can change one of these arguments to \texttt{False}.

The method \texttt{display} shows the Christoffel components in the output of the cell. This method receives a single string with a sequence of symbols $\,\hat{ }\,$ or $\_$ separated with commas. For example, if we want to display the Christoffel symbol of second kind $\Gamma {}^\gamma{}_\alpha{}_\beta$ we call the method as \texttt{Chr.display("\,\,$\hat{ }$\,\,,\_,\_")}. Note that the \texttt{display} method only shows the non-zero components.

If the user tries to display other combinations of indices, the program will raise an error because there are only two possible combinations: first and second. We decide to leave the \texttt{display} method this way, to have a consistent notation regarding the methods of a Tensor in Pytearcat. It also allows to directly understand which combination of indices are being displayed directly from the line of code.

\subsubsection{Riemann tensor}

The Riemann tensor is defined as

\begin{equation}
{R^{\rho }}_{{\sigma \mu \nu }}=\partial _{\mu }\Gamma _{{\nu \sigma }}^{\rho }-\partial _{\nu }\Gamma _{{\mu \sigma }}^{\rho }+\Gamma _{{\mu \lambda }}^{\rho }\Gamma _{{\nu \sigma }}^{\lambda }-\Gamma _{{\nu \lambda }}^{\rho }\Gamma _{{\mu \sigma }}^{\lambda }.
\label{eq:Riemann}
\end{equation}

\begin{figure}
    \centering
    \includegraphics[width=0.8\textwidth]{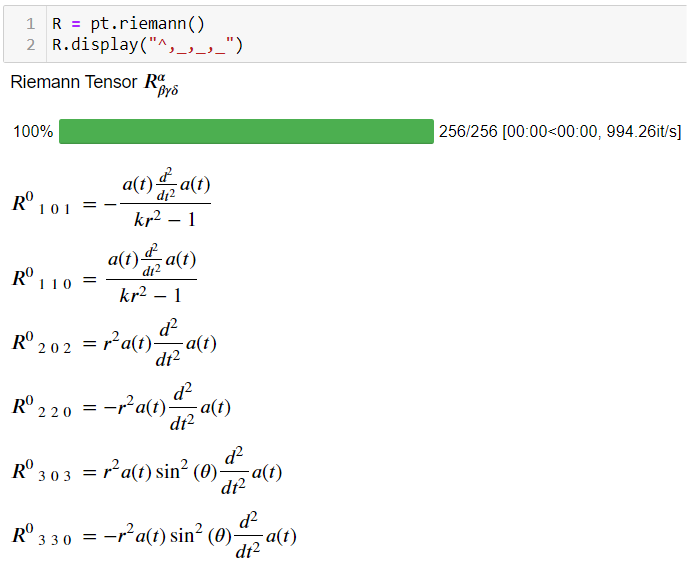}
    \caption{Example of the calculation of the Riemann Tensor, together with the respective output. In this screenshot we do not show all the Riemann components.}
    \label{fig:Riemann}
\end{figure}

In Pytearcat, we calculate the Riemann tensor calling the function \texttt{riemann}, which will return a \texttt{Tensor} object. This function, by default, calculates the Riemann tensor with the first index contravariant and the others covariant, as expressed in equation \eqref{eq:Riemann}. We can specify if we want to calculate all the indices combinations or only the default indices combination. The argument to do this is \texttt{All = True}. The function is shown in Figure \ref{fig:Riemann}, as well as the corresponding output.

\subsubsection{Ricci tensor and Ricci scalar}

The Ricci tensor is defined as

\begin{equation}
{\displaystyle R_{\sigma \nu }={R^{\rho }}_{\sigma \rho \nu }=\partial _{\rho }\Gamma _{\nu \sigma }^{\rho }-\partial _{\nu }\Gamma _{\rho \sigma }^{\rho }+\Gamma _{\rho \lambda }^{\rho }\Gamma _{\nu \sigma }^{\lambda }-\Gamma _{\nu \lambda }^{\rho }\Gamma _{\rho \sigma }^{\lambda }},
\label{eq:RicciTensor}
\end{equation}
and the Ricci scalar is given by

\begin{equation}
S  = g^{\sigma\nu}R_{\sigma\nu} = R^\nu_\nu.
\end{equation}

\begin{figure}
    \centering
    \includegraphics[width=0.8\textwidth]{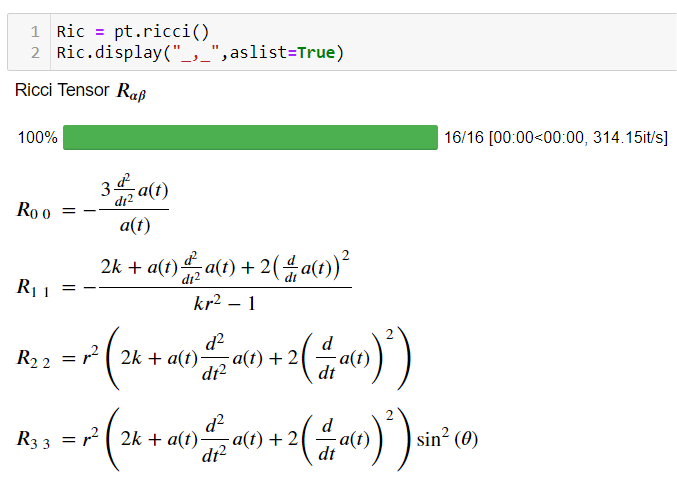}
    \caption{Example of the calculation of the Ricci Tensor, together with the respective output.}
    \label{fig:ricci_tensor}
\end{figure}

We calculate the Ricci tensor in Pytearcat by calling the function \texttt{ricci}, which will return a \texttt{Tensor} object. By default, this function calculates the Ricci tensor with the two covariant indices as in equation \eqref{eq:RicciTensor}. The function called and its respective output are shown in Figure \ref{fig:ricci_tensor}. Since the Ricci tensor is a 2-rank tensor, the \texttt{display} method in Pytearcat will show the Ricci tensor in a matrix form, unless the method receives the argument \texttt{aslist=True} as in Figure \ref{fig:ricci_tensor}. 

\begin{figure}
    \centering
    \includegraphics[width=0.8\textwidth, height=3.4cm]{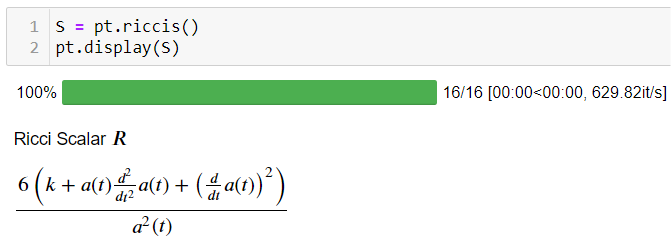}
    \caption{Example of the calculation of the Ricci scalar with its corresponding output.}
    \label{fig:Ricci_scalar}
\end{figure}

Similarly, the Ricci scalar is calculated by calling the function \texttt{riccis}, but this function receives no argument because it returns a scalar (a non-tensor object). By default this will be a sympy object. This function is  shown in Figure \ref{fig:Ricci_scalar}. Note that in this example, we used the \texttt{display} function and not the method because the returned object is a Sympy object, thus it does not have the display method defined for tensor objects in Pytearcat.

\subsubsection{Einstein tensor}

The Einstein tensor is given by

\begin{equation}
G_{\mu \nu }=R_{\mu \nu }-{\frac{1}{2}}g_{\mu \nu }R.
\label{eq:einstein}
\end{equation}

\begin{figure}
    \centering
    \includegraphics[width=0.8\textwidth]{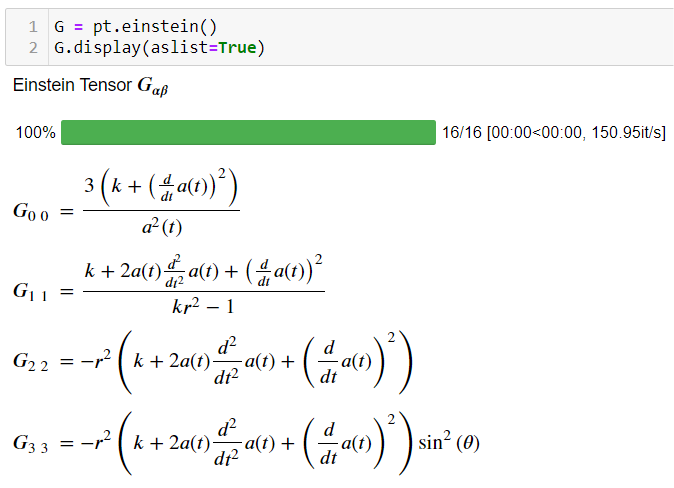}
    \caption{Calculation of the Einstein Tensor, together with the respective output.}
    \label{fig:einstein}
\end{figure}

In Pytearcat, to calculate the Einstein tensor, we call the function \texttt{einstein}. This will return a \texttt{Tensor} object. By default, this function calculates the Einstein tensor with the two covariant indices as defined in equation \eqref{eq:einstein}. This function is shown in Figure \ref{fig:einstein}, where we included the optional argument \texttt{aslist=True} to the \texttt{display} method since again, this is a 2nd-rank tensor.

 Finally, note that if we call a function to calculate some of these quantities without calling the other instances before (for example, if we ask Pytearcat to calculate the Einstein tensor without calling before the Christoffel symbols), the code will automatically calculate all the necessary objects. Furthermore, if a \texttt{Tensor} has been calculated before, the program will alert the user, printing a message that indicates which tensor object has already been calculated.

\section{Tensors}\label{sec:tensor}

As explained in Sect. \ref{sec:GR-Tensor}, \texttt{Tensor} objects in Pytearcat are flexible enough to define and operate with general tensors, following the usual rules of raising and lowering indices. In this section, we will show how to define arbitrary tensors and assign values to their components. We also explain how to operate tensor objects, including the Einstein summation and index contraction, and present useful routines available on Pytearcat. 

\subsection{Creating a \texttt{Tensor}}

In Pytearcat, we can define arbitrary tensors using the function \texttt{ten}. For example, a tensor of rank $n$ and named $A$ can be created writing \texttt{pt.ten("A",n)}. We also can assign values to the components of the tensor using the \texttt{Tensor} method \texttt{assign}, indicating the values to be assigned on a specific indices combination. We can raise and lower indices and operate over the indices applying the Einstein summation. Pytearcat has been designed to operate as is usual in physics. It is important to note that a \texttt{Tensor} object has many attributes shown in Table \ref{tab:attributes}.

\begin{table}
    \centering
    \caption{Attributes of a \texttt{Tensor} object in Pytearcat.}
    \begin{tabular}{p{1.5cm} p{6.5cm}}
    \hline
        Attribute & Description \\ \hline
        \texttt{indices} & List of Boolean values indicating which tensor indices combination is already calculated.\\
        \texttt{n} & Rank of the tensor.\\
        \texttt{name} & Name of the tensor object.\\
        \texttt{sequence} & List of strings to indicate all the possible components. For example, a 2-rank tensor will have the following associated list: \text{['\_,\_', '\_,$^\wedge$', '$^\wedge$ ,\_', '$^\wedge$ , $^\wedge$ ']}\\
        \texttt{tensor} & Nested list that contains the data of each component of the tensor.\\
        \hline
    \end{tabular}
    \label{tab:attributes}
\end{table}

\begin{figure}
    \centering
    \includegraphics[width=0.7\textwidth]{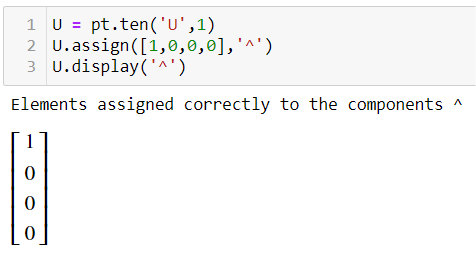}
    \caption{Definition of the 4-velocity in Pytearcat.}
    \label{fig:4velocity}
\end{figure}

\begin{figure}
    \centering
    \includegraphics[width=0.4\textwidth]{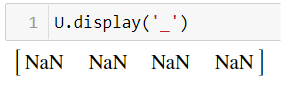}
    \caption{Output of the display method on the covariant 4-velocity.}
    \label{fig:4velocityundefoutput}
\end{figure}

By default every instance of \texttt{Tensor} has all the elements of each component defined as \texttt{NaN} (\textit{Not a Number}). If we want to work with some tensors or elements with indices, first of all we have to define the elements. In order to show how the user can define and work with other tensors in Pytearcat, we will show an example defining the energy-momentum tensor. This tensor is denoted as $T{}_\mu{}_\nu$ and is given by:
 
\begin{equation}
T{}^\mu{}^\nu = \left(\rho+P\right)U{}^\mu U{}^\nu+P g{}^\mu{}^\nu,
\label{eq:energy-momentum}
\end{equation}

where we used $c=1$. Before defining $T{}^\mu{}^\nu$, we have to define the 4-velocity $U{}^\mu$. In Pytearcat we define this object as is shown in Figure \ref{fig:4velocity}.
If the user tries to display the covariant 4-velocity, the program will show a horizontal vector filled with \texttt{NaN} values because this component has not been assigned (see Figure \ref{fig:4velocityundefoutput}).

\begin{figure}
    \centering
    \includegraphics[width=0.7\textwidth]{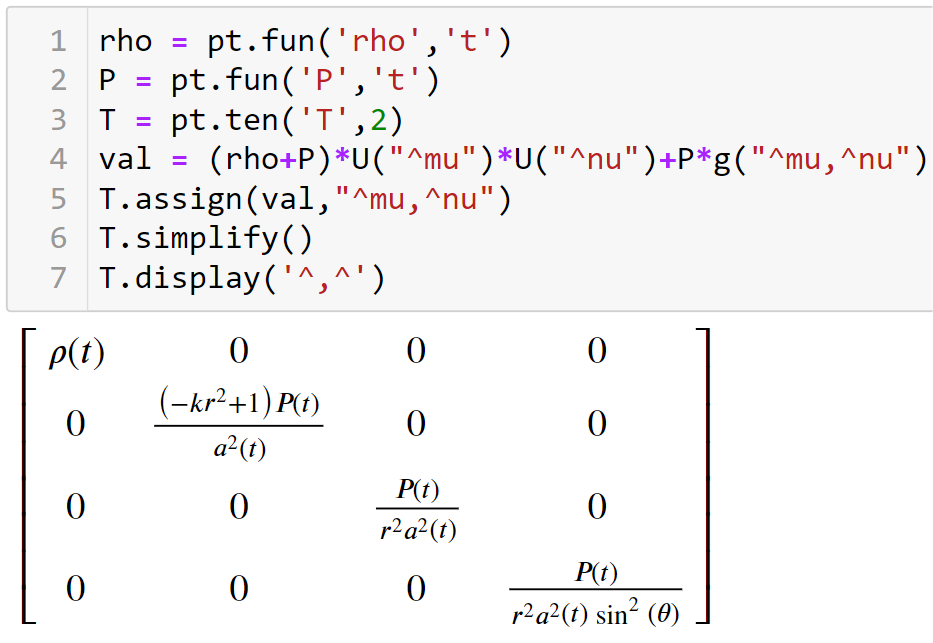}
    \caption{This is an example of a definition and assignation of a \texttt{Tensor}. In this case, we define the functions $\rho(t)$ and $P(t)$ to calculate $T^{\mu\nu}$, compute the value and assign it to the $\texttt{Tensor}$ $T{}^\mu{}^\nu$.}
    \label{fig:Tmunu asignation}
\end{figure}

Now we define the functions $\rho(t)$ and $P(t)$, and then we also define the 2-rank tensor as $T$. Finally, we assign the result of the operation given by equation \eqref{eq:energy-momentum} as a component of the $T{}^\mu{}^\nu$. This assignation is made with the method \texttt{assign}. The user has to give a second argument that indicates to which component of the tensor $T$ wants to assign the result of this mathematical operation. This second argument has to be a string, as shown in Figure \ref{fig:Tmunu asignation}.

 We stress that the user can assign values by hand to all the indices combinations, and this object will never have values according to the transformation rules. This is not a problem, but the user has to know what he wants to define. For instance, the user can define a Christoffel symbol by hand. In this case, the \texttt{Tensor} objects will contain all the indices combinations ($\Gamma^{\alpha \beta \gamma}$, $\Gamma^{\alpha \beta}{}_{\gamma}$, ... ,$\Gamma_{\alpha \beta \gamma}$), but these objects are not real tensors, because they do not follow the transformation rules
\begin{equation}
    {\displaystyle {\hat {T}}_{j'_{1}\dots j'_{q}}^{i'_{1}\dots i'_{p}}={\frac {\partial {\bar {x}}^{i'_{1}}}{\partial x^{i_{1}}}}\cdots {\frac {\partial {\bar {x}}^{i'_{p}}}{\partial x^{i_{p}}}}{\frac {\partial x^{j_{1}}}{\partial {\bar {x}}^{j'_{1}}}}\cdots {\frac {\partial x^{j_{q}}}{\partial {\bar {x}}^{j'_{q}}}}T_{j_{1}\dots j_{q}}^{i_{1}\dots i_{p}}.}
\end{equation}
 However, in the context of Pytearcat they belong to the \texttt{Tensor} class.
 
\begin{figure}
    \centering
    \includegraphics[width=0.9\textwidth]{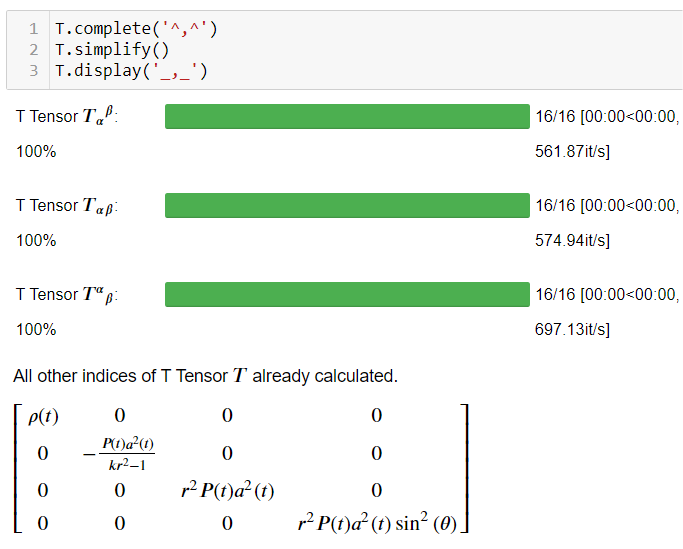}
    \caption{Example of the \texttt{complete} method.}
    \label{fig:Tmunu complete}
\end{figure}

To calculate all the other components of the $T$ tensor, we can use the method \texttt{complete}. This receives a string that indicates the indices combination from which Pytearcat will automatically raise and lower the indices to complete the \texttt{Tensor}. An example of this method appears on Figure \ref{fig:Tmunu complete}.

At this point, we can find the conservation equations of the $T^{\mu\nu}$ tensor

\begin{equation}
    \nabla_\nu T^{\mu\nu} = 0.
    \label{eq:Tmunu conservation}
\end{equation}
In Pytearcat, the user can calculate the partial derivative $\partial_\nu$ by using the function \texttt{D}. Additionally, the covariant derivative $\nabla_\nu$ is calculated using the function \texttt{C}. The user could be familiar with the following notations

\begin{align}
&\partial^\nu \equiv g{}^\mu{}^\nu\partial_\mu \equiv g{}^\mu{}^\nu\frac{\partial}{\partial x^\mu}, \label{eq:partial_up}\\
&\nabla^\nu \equiv g{}^\mu{}^\nu\nabla_\mu.\label{eq:covariant_up}
\end{align}

In Pytearcat, it is impossible to derive with respect to an upper index (in both partial and covariant derivatives). However, the user can explicitly calculate the right-hand side of Eqns. \eqref{eq:partial_up} and \eqref{eq:covariant_up} to obtain the desired result.

\begin{figure}
    \centering
    \includegraphics[width=0.5\textwidth]{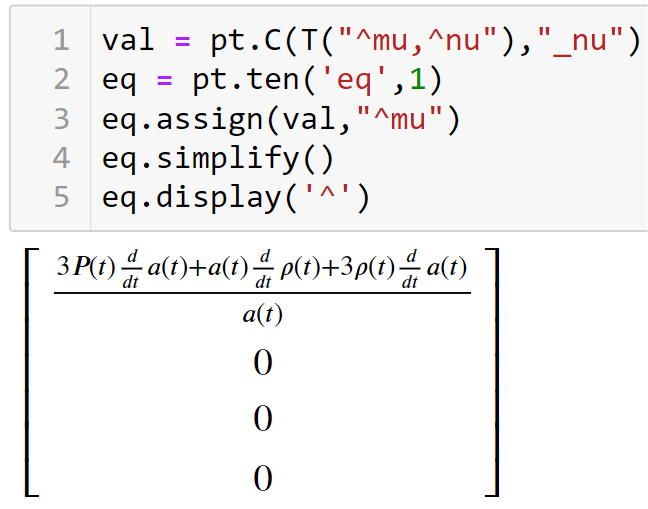}
    \caption{This is an example that shows how the user can calculate the conservation of the tensor $T$ through the covariant derivative.}
    \label{fig:conservationT}
\end{figure}

To calculate the conservation equations for the energy-momentum tensor (equation \eqref{eq:Tmunu conservation}), the user calculates the covariant derivative and assign its output to a variable named \texttt{val} (Figure \ref{fig:conservationT}). Then, a new \texttt{Tensor} is created and the variable is assigned to the desired index configuration of the \texttt{Tensor}. Finally the resulting conservation equations are shown using the \texttt{display} method (Figure \ref{fig:conservationT}). In Pytearcat the Einstein Summation is applied automatically. 

In order to clarify the operations executed by \texttt{Pytearcat}, in this example $T$ is a \texttt{Tensor} (an instance of the \texttt{Tensor} class), and \texttt{T("\_mu,\_nu")}is a \texttt{Tdata} object (an instance of the \texttt{Tdata} class) that contains all the information of $T_{\mu \nu}$ with $\mu,\nu = 0,1,2,3$. In general the information of a \texttt{Tensor} with rank $n$ is characterised by $2^n$ \texttt{Tdata} objects. 

In Pytearcat we are operating \texttt{Tdata} objects and not \texttt{Tensor} objects, because the indexation of the \texttt{Tensor} object is important. This indexation is made through the built-in method \texttt{call}, e.g. \texttt{T("\_mu,\_nu")}. The user can operate \texttt{Tdata} objects as in physics; in other words, the Einstein summation is automatically executed in these expressions. The Einstein summation also acts where there are partial derivatives or covariant derivatives. In the following subsections, we show other examples where the Einstein summation and the derivatives are applied.

\subsection{Einstein summation}

In order to show an example of the automatic Einstein summation in Pytearcat, we work with the electromagnetic tensor. First of all, we have to define a Minkowski space-time given by 

\begin{equation}
ds^2 = dt^2 - dx^2 -dy^2 - dz^2,
\label{minkowskimetric}
\end{equation}
and also we work with the electromagnetic 4-potential described by

\begin{equation}
A^{\alpha }=(\phi ,\mathbf {A} ),
\end{equation}
where $\phi$ is the electric potential and $A$ is the vector potential. In physics, the electric field $\mathbf{E}$ and the magnetic field $\mathbf{B}$ are determined by

\begin{align}
&\mathbf {E} =-\mathbf {\nabla } \phi -{\frac {\partial \mathbf {A} }{\partial t}}, \label{eq:E}\\
&{\displaystyle \mathbf {B} =\mathbf {\nabla } \times \mathbf {A} }. \label{eq:B}
\end{align}

With these definitions, the electromagnetic tensor $F{}^\mu{}^\nu$ is defined as

\begin{equation}
{\displaystyle F^{\mu \nu }=\partial ^{\mu }A^{\nu }-\partial ^{\nu }A^{\mu }={\begin{bmatrix}0&-E_{x}/c&-E_{y}/c&-E_{z}/c\\E_{x}/c&0&-B_{z}&B_{y}\\E_{y}/c&B_{z}&0&-B_{x}\\E_{z}/c&-B_{y}&B_{x}&0\end{bmatrix}}}.
\label{eq:Fmunu}
\end{equation}

\begin{figure}
    \centering
    \includegraphics[width=0.8\textwidth]{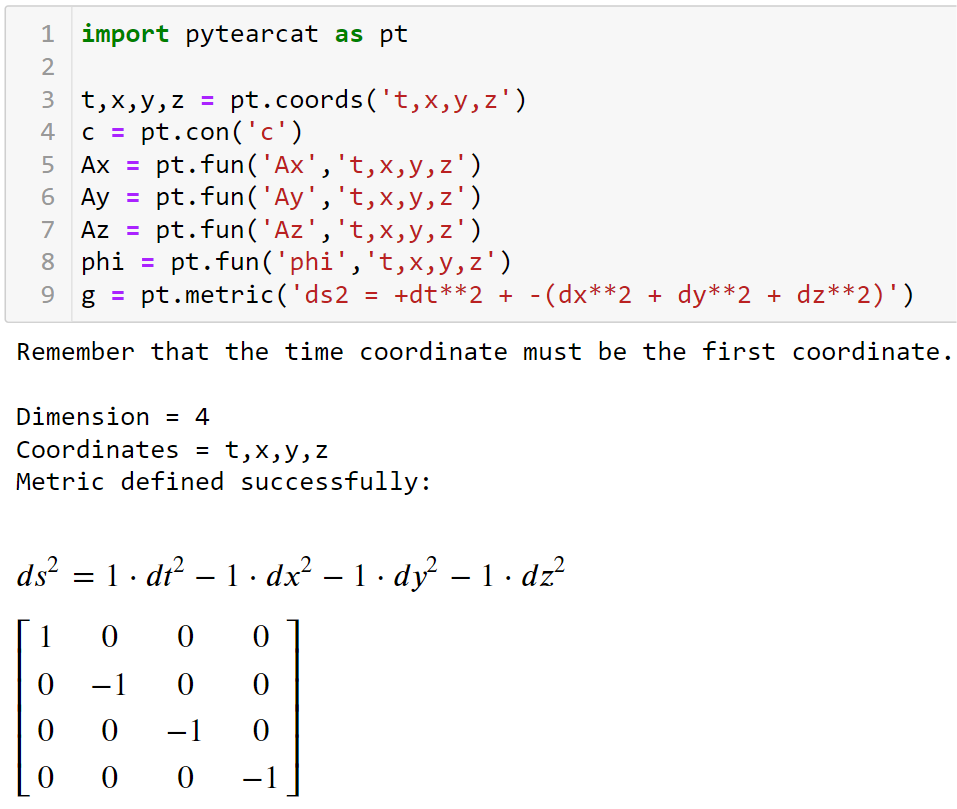}
    \caption{This example shows how to define the Minkowski metric, defining the coordinates, and the potentials $\mathbf{A}$ and $\phi$, and finally calling the function \texttt{metric} and passing it as an argument a string with the line element.}
    \label{fig:minkowski}
\end{figure}

\begin{figure}
    \centering
    \includegraphics[width=0.9\textwidth]{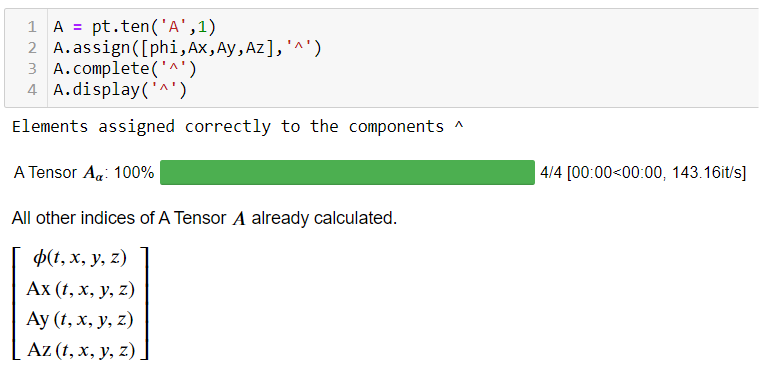}
    \caption{Example of the definition of the electromagnetic 4-potential in Pytearcat.}
    \label{fig:4potential}
\end{figure}

\begin{figure}
    \centering
    \includegraphics[width=0.7\textwidth]{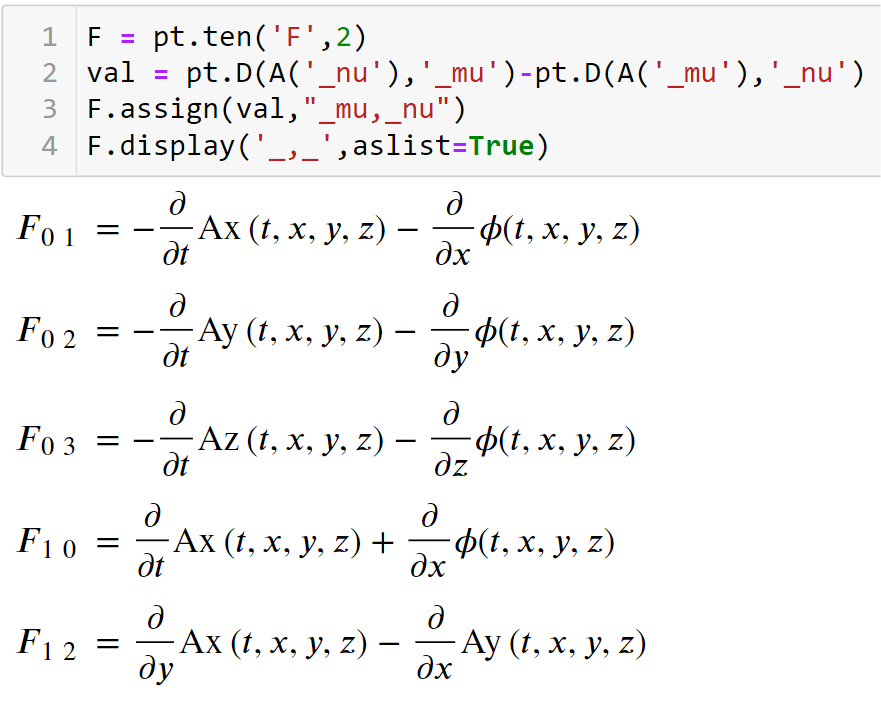}
    \caption{Example of the definition of the electromagnetic tensor $F$ in Pytearcat. In this figure we do not show all the components of $F_{\mu\nu}$.}
    \label{fig:Fmunu}
\end{figure}

In Pytearcat we define the Minkowski metric and the functions in Figure \ref{fig:minkowski}.
The electromagnetic 4-potential is defined through the \texttt{assign} method of the tensor $A$ that we create in Figure \ref{fig:4potential}.
With this, we can calculate the electromagnetic tensor and use the \texttt{assign} method to store it in the \texttt{Tensor} $F$ defined in Figure \ref{fig:Fmunu}.

\begin{figure}
    \centering
    \includegraphics[width=0.9\textwidth]{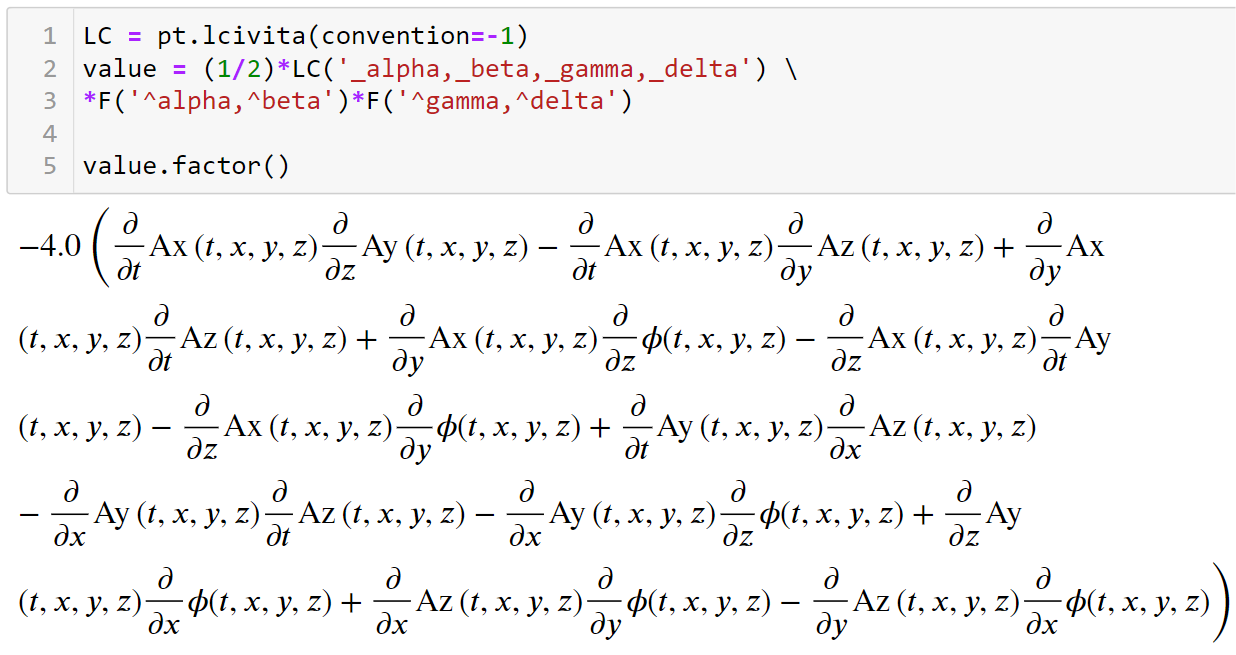}
    \caption{Example of the \textit{Pseudo-scalar} invariant in Pytearcat.}
    \label{fig:pseudoescalarinvariant}
\end{figure}

With these definitions, the user can operate the Tensor components efficiently. For example, there is a quantity known as the \textit{pseudoscalar invariant} which is defined as

\begin{equation}
{\frac {1}{2}}\epsilon _{\alpha \beta \gamma \delta }F^{\alpha \beta }F^{\gamma \delta }=-4\mathbf {B} \cdot \mathbf {E},
\label{eq:pseudoscalarinvariant}
\end{equation}
where we have used $c = 1$ and $\epsilon _{\alpha \beta \gamma \delta }$ is the Levi-Civita symbol. Note that  the Levi-Civita symbol is defined as

\begin{equation}
\varepsilon _{a_{1}a_{2}a_{3}\ldots a_{n}}={\begin{cases}+1&{\text{if }}{\text{is an even permutation,}}\\
-1&{\text{if is an odd permutation, }}\\
\;\;\,0&{\text{otherwise}},\end{cases}}
\label{eq:levicivitasymbol}
\end{equation}
where $(a_{1},a_{2},a_{3},\ldots ,a_{n})$ is a permutation of $(1,2,3,\ldots ,n)$. In this example, we use the following convention:

\begin{equation}
\epsilon _{0123}=-1.
\end{equation}
This convention can be declared calling the \texttt{lcivita} object and including the argument \texttt{``convention = -1''} (-1 indicates that $\epsilon _{0123}=-1$. By default, \texttt{``convention''} is set to be equal to 1 which implies $\epsilon _{0123}=1$). In Fig \ref{fig:pseudoescalarinvariant} we show an example of the definition of the Levi-Civita symbol with the desired convention. In Pytearcat, the Levi-Civita symbol contains information about the permutations, and it is not a physical tensor. Even this symbol has no meaning under raising and lowering indices. In Pytearcat, this definition is exactly equal for any component of the Levi-Civita tensor. For example, $\epsilon_{0123}=\epsilon_{0}{}^1{}_2{}^3=\epsilon^{0123}$. The Levi-Civita symbol in Pytearcat does not consider if the indices are contravariant or covariant. The code and the output are shown in Figure \ref{fig:pseudoescalarinvariant}. Note that the Einstein summation given by equation \eqref{eq:pseudoscalarinvariant} runs over four indices automatically.

Considering the definitions given by the Eqns. \eqref{eq:E} and \eqref{eq:B}, the output shown in Figure \ref{fig:pseudoescalarinvariant} corresponds to $-4\mathbf {B} \cdot \mathbf {E}$, which agrees with the definition on equation \eqref{eq:pseudoscalarinvariant}.

\begin{figure}
    \centering
    \includegraphics[width=0.25\textwidth]{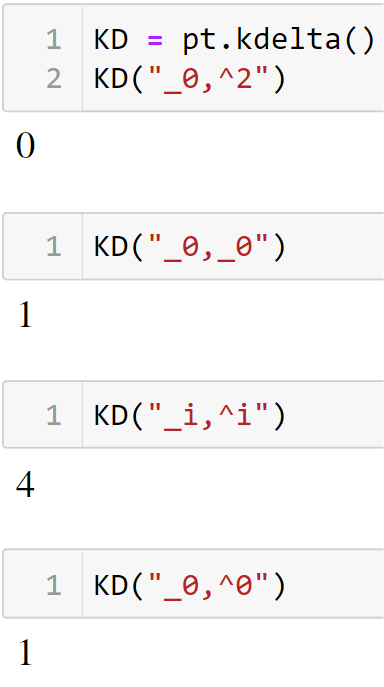}
    \caption{Example of the Kronecker Delta in Pytearcat.}
    \label{fig:deltaK}
\end{figure}

In addition to the Levi-Civita symbol, Pytearcat also has another helpful symbol, the Kronecker Delta. This symbol is defined as

\begin{equation}
\delta _{{ij}}\equiv\delta_i{}^j\equiv\delta ^j{}_i\equiv\delta ^{{ij}}\equiv{\begin{cases}0&{\text{if }}i\neq j,\\1&{\text{if }}i=j,\end{cases}}
\label{eq:deltakronecker}
\end{equation}
and can be defined as shown in Fig \ref{fig:deltaK}. Note that in this figure, we can see that the Einstein Summation runs automatically if the the indices are repeated letters but if the indices are integers, then it returns the corresponding elements of that indices, i.e, $\delta_0{}^2$, $\delta_{0}{}_{0}$ and $\delta_0{}^0$ for the first, second and fourth executions of the method \texttt{call} in Figure \ref{fig:deltaK}.

\subsection{Summing only over space}

\begin{figure}
    \centering
    \includegraphics[width=0.6\textwidth]{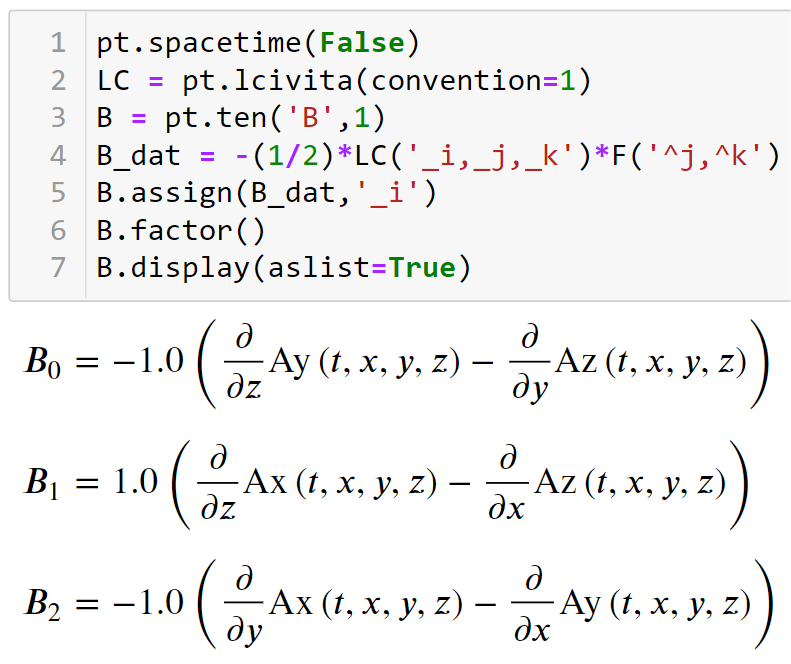}
    \caption{Example of the Einstein summation running only on the space coordinates.}
    \label{fig:sum_over_space}
\end{figure}

The automatic Einstein summation is programmed to sum over all the repeated indices over all the dimensions. If the user wants to sum only over the indices associated with space, then the user has to call the function \texttt{spacetime} where the argument is a Boolean. If it is \texttt{True} (as default), then Pytearcat sums over all the indices, but if the argument is \texttt{False}, then the sum considers only the indices associated with space. Figure \ref{fig:sum_over_space} shows an example of this, where we work with the spatial part of the $F^{\mu\nu}$ tensor and with the Levi-Civita symbol to obtain the definition of the magnetic field. In physics, the relationship between $\mathbf{B}$ and $F^{\mu\nu}$ is given by

\begin{equation}
B_{i}=-{\frac {1}{2}}\epsilon _{ijk}F^{jk},
\label{eq:B_and_F}
\end{equation}
summing over the spatial part. In this code we have used the Levi-Civita symbol defined with the usual convention, and also we used the method \texttt{factor} to factorise the components. To go back to the standard Einstein summation (over the complete space-time), the user must call the function \texttt{spacetime} with the argument \texttt{True}.

\subsection{Series expansion}

Pytearcat can also work using series and approximation up to order $\mathcal{O}(n)$. This feature is handy in cases such as perturbation theory \citep{book:Piattella:2018}. In this scenario, the calculations can be improved in time because the amount of mathematical operations decreases.

\begin{figure}
    \centering
    \includegraphics[width=\textwidth]{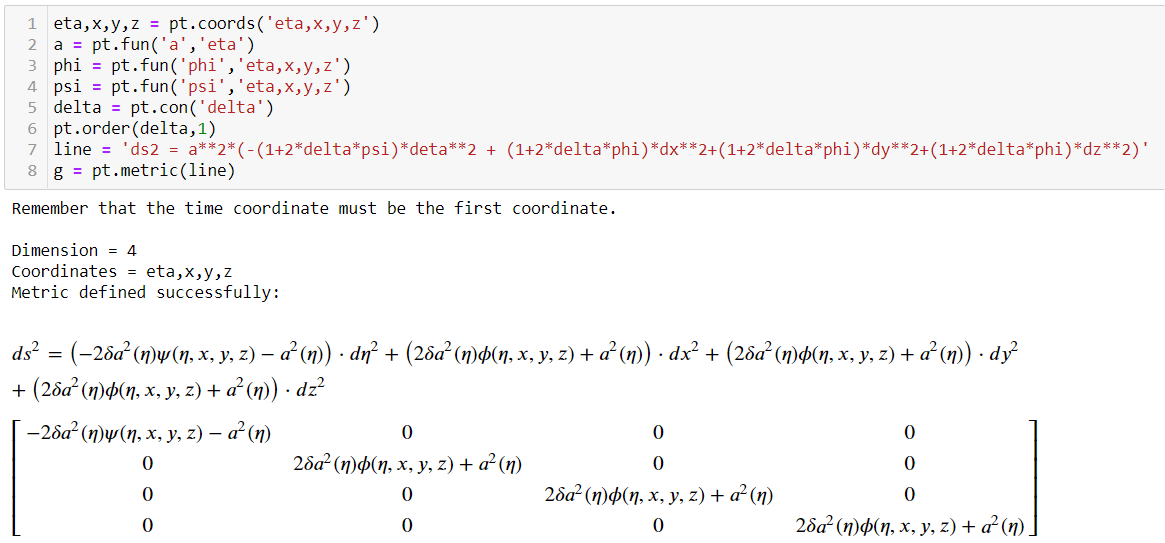}
    \caption{An example of series in cosmological perturbation theory only with the scalar contribution.}
    \label{fig:perturbative_metric}
\end{figure}

\begin{figure}
    \centering
    \includegraphics[width=0.8\textwidth]{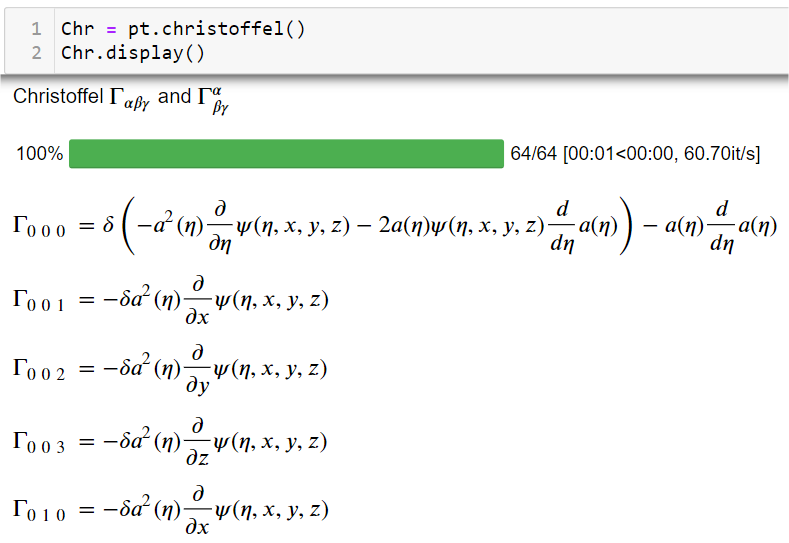}
    \caption{An example of Christoffel symbols calculated using series expansion. In this figure we do not show all the components of the Christoffel symbols.}
    \label{fig:christoffel_series}
\end{figure}

This feature is activated when the user calls the function \texttt{order}. It has two arguments: the variable where the series expansion will occur and the maximum expansion order the program preserves in the calculations. For instance, if the user declare \texttt{order(delta,1)}, the program will preserve all the terms up to first order in the variable \texttt{delta}. We include an example of the cosmological perturbation theory in the scalar sector, where $\phi$ and $\psi$ are potentials (first-order). This characteristic is represented using an extra parameter ($\delta$). In Figure \ref{fig:perturbative_metric} we define the FLRW metric in perturbation theory, including the potentials and using a series expansion up to $\mathcal{O(\delta)}$. Note that \texttt{delta} must be a constant and must be defined before calling \texttt{setorder}. From this declaration, every calculation will consider the order expansion of $\delta$. For example, in Figure \ref{fig:christoffel_series} the Christoffel symbols are automatically calculated considering the series expansion.

The constant use as the expansion variable can be declared with an arbitrary name. However, the name \texttt{epsilon} is forbidden because of name conflicts with some of the core modules in Pytearcat. We recommend the use of greek letters because the method \texttt{display} will render the letter giving a much better experience.

\section{Execution times}\label{sec:execution}

One crucial aspect of any computational tool is the execution time of the code itself. However, in tensor calculus, the user will execute different calculations with different complexities, implying that the execution time will depend strictly on the required task. To give some estimations, we prepare a typical workload for the package starting from the definition of the metric and ending with the calculation of the Einstein Tensor, $G$. 

As python is an inefficient programming language, many modules and implementations are built to run in more efficient languages such as C. For this reason, as explained in Sec. \ref{sec:structure}, Pytearcat can work with Sympy or Giacpy as the core of the symbolic calculations. Also, other symbolic tensor calculators are available for other platforms. We compare Pytearcat with GRtensorIII working on Maple to present a broader view of the capabilities of Pytearcat. 

The test consists of calculating the Christoffel symbols, the Riemann tensor, the Ricci tensor, the Ricci scalar, and the Einstein tensor. The test uses the following metric 

\begin{equation}
\begin{split}
    \displaystyle ds^2 = &\left(-1-2\,\phi\left(t,x,y,z\right) \delta\right) \cdot dt^2
    +\left(a\left(t\right)^{2}-2\,a\left(t\right)^{2}\,\psi\left(t,x,y,z\right) \delta\right) \cdot dx^2\\
    & +\left(a\left(t\right)^{2}-2\,a\left(t\right)^{2}\,\psi\left(t,x,y,z\right) \delta\right) \cdot dy^2
     +\left(a\left(t\right)^{2}-2\,a\left(t\right)^{2}\,\psi\left(t,x,y,z\right) \delta\right) \cdot dz^2,
\end{split}
\label{eq: Performance Test Metric}
\end{equation}
running on an Intel Core I7-8550U CPU at 1.80 GHz. The $\delta$ variable in the metric corresponds to the series expansion variable implying that during the execution, the program expands around this variable.

\begin{table}
    \centering
    \caption{Execution times for different symbolic calculators.}
    \begin{tabular}{l c c c}
    \hline 
        Task & \multicolumn{3}{c}{Execution time (ms)} \\
         & Pytearcat (Sympy) & Pytearcat (Giacpy) & Maple (GRtensor)\\ \hline
         Christoffel & $2640 \pm 537$ & $499 \pm 34$ & $9 \pm 8$\\
     Riemann & $4980 \pm 611$ & $1280 \pm 82$ & $60 \pm 10$\\
     Ricci Tensor& $488 \pm 39$ & $138 \pm 6$ & $11 \pm 8$\\
     Ricci Scalar & $153 \pm 12$ & $62 \pm 3$ & $5 \pm 8$\\
     Einstein & $452 \pm 19$ & $115 \pm 4$ & $11 \pm 13$\\\hline
     \textbf{Total} & $8713 \pm 1218$ & $2094 \pm 128$ & $95 \pm 18$\\ \hline
    \end{tabular}
    \label{tab:Execution Times}
\end{table}

Table \ref{tab:Execution Times} summarises the execution times of the different tasks for the three different symbolic calculators. Pytearcat running with Giacpy is up to 4 times faster than running with Sympy. GRtensorIII running on Maple is up to 20 times faster than Pytearcat with Giacpy. We stress that all the tasks are completed in seconds.

\section{Conclusions}

Pytearcat is a tensor algebra calculator created to be used within Jupyter Notebooks (Python) as a versatile and intuitive tool for the user. Pytearcat is built around the GR framework, following notations and definitions as used in physics and astrophysics. The greatest strengths of Pytearcat are the similarity of its syntax with the physics notation and the Einstein notation ($g_{\alpha\beta} T^{\alpha \gamma}$), the evaluation of a tensor component ($T_{0}{}_1$) and even mixed cases ($g^{\alpha \beta}T_{\alpha 0}$), where the user can create arbitrary tensors. Also, the user can choose to work with the space-time components or only with the space components, which works with the Einstein summation. Other advantages include that the metric can be redefined in the same notebook without restarting the Kernel and the possibility of using series expansions. Furthermore, all the outputs can be easily exported to \LaTeX $\, $(secondary button on the rendered output). 

As the whole purpose of using computers to perform tensor calculus is to do it fast, we put much effort in optimising the calculations, employing tensor symmetries, and using faster symbolic calculator libraries. We also compared the execution times for different symbolic calculators, finding that Pytearcat running with Giacpy is about four times faster than Pytearcat running with Sympy but is still slower than other calculators. However, it is worth pointing out that even if Pytearcat does not overcome the speed of some calculator such as GRtensor, it is free to use and can be improved in collaboration with the community (open-source).

This is the first release of Pytearcat, and we expect to include more features in the package. Some of them include implementing symbolic calculations on differential geometry with a more mathematical notation and other features to work with Lagrangians using index notation (Quantum Field Theory) and space-times with torsion.

\section*{Acknowledgements}

Funding: This work was supported by CONICYT project Basal [grant number AFB-170002]; and Fondecyt Regular [Grant Number 1191813]

\bibliographystyle{mnras}

\bibliography{references}

\begin{thebibliography}{}
\makeatletter
\relax
\def\mn@urlcharsother{\let\do\@makeother \do\$\do\&\do\#\do\^\do\_\do\%\do\~}
\def\mn@doi{\begingroup\mn@urlcharsother \@ifnextchar [ {\mn@doi@}
  {\mn@doi@[]}}
\def\mn@doi@[#1]#2{\def\@tempa{#1}\ifx\@tempa\@empty \href
  {http://dx.doi.org/#2} {doi:#2}\else \href {http://dx.doi.org/#2} {#1}\fi
  \endgroup}
\def\mn@eprint#1#2{\mn@eprint@#1:#2::\@nil}
\def\mn@eprint@arXiv#1{\href {http://arxiv.org/abs/#1} {{\tt arXiv:#1}}}
\def\mn@eprint@dblp#1{\href {http://dblp.uni-trier.de/rec/bibtex/#1.xml}
  {dblp:#1}}
\def\mn@eprint@#1:#2:#3:#4\@nil{\def\@tempa {#1}\def\@tempb {#2}\def\@tempc
  {#3}\ifx \@tempc \@empty \let \@tempc \@tempb \let \@tempb \@tempa \fi \ifx
  \@tempb \@empty \def\@tempb {arXiv}\fi \@ifundefined
  {mn@eprint@\@tempb}{\@tempb:\@tempc}{\expandafter \expandafter \csname
  mn@eprint@\@tempb\endcsname \expandafter{\@tempc}}}

\bibitem[\protect\citeauthoryear{{Bapat} et~al.,}{{Bapat}
  et~al.}{2020}]{Einsteinpy:2020}
{Bapat} S.,  et~al., 2020, {EinsteinPy: General Relativity and gravitational
  physics problems solver} (\mn@eprint {ascl} {2012.026})

\bibitem[\protect\citeauthoryear{Carroll}{Carroll}{2014}]{book:Carroll:2014}
Carroll S.~M.,  2014, Spacetime and geometry: an introduction to general
  relativity, pearson new international edition edn.
Always learning, Pearson Education. C 2014

\bibitem[\protect\citeauthoryear{Meurer et~al.,}{Meurer et~al.}{2017}]{sympy}
Meurer A.,  et~al., 2017, \mn@doi [PeerJ Computer Science]
  {10.7717/peerj-cs.103}, 3, e103

\bibitem[\protect\citeauthoryear{{Piattella}}{{Piattella}}{2018}]{book:Piattella:2018}
{Piattella} O.,  2018, {Lecture Notes in Cosmology}.
Springer International Publishing, \mn@doi{10.1007/978-3-319-95570-4}

\bibitem[\protect\citeauthoryear{{San Mart{\'\i}n}, {Alfaro}  \& {Rubio}}{{San
  Mart{\'\i}n} et~al.}{2021}]{SanMartin:2021}
{San Mart{\'\i}n} M.,  {Alfaro} J.,   {Rubio} C.,  2021, \mn@doi [The
  Astrophysical Journal] {10.3847/1538-4357/abddc3}, \href
  {https://ui.adsabs.harvard.edu/abs/2021ApJ...910...43S} {910, 43}

\makeatother
\end{thebibliography}


\end{document}